\documentclass[10pt]{article}
\usepackage{amssymb, amsmath}
\usepackage[dvipdfm]{graphicx}
\begin{document}

\makeatletter
%%%%%------- preprint style ----%%%%%%%%%%%%%%%%%%%%
\@addtoreset{equation}{section}
\def\theequation{\thesection.\arabic{equation}}
\def\@maketitle{\newpage
 \null
 {\normalsize \tt \begin{flushright} 
  \begin{tabular}[t]{l} \@date  
  \end{tabular}
 \end{flushright}}
 \begin{center} 
 \vskip 2em
 {\LARGE \@title \par} \vskip 1.5em {\large \lineskip .5em \begin{tabular}[t]{c}\@author 
 \end{tabular}\par} 
 \end{center}
 \par
 \vskip 1.5em} 
\makeatother
%%%%%%%%%%%%%%%%%%%%%%%%%
\topmargin=-1cm
\oddsidemargin=1.5cm
\evensidemargin=-.0cm
\textwidth=15.5cm
\textheight=22cm
%\renewcommand{\baselinestretch}{1.5}
%%%%%%%%%%%%%%%%%%%%%%%%
\setlength{\baselineskip}{16pt}
\title{ BMS$_4$ Surface-Charge Algebra \\ via Hamiltonian Framework}
\author{
Ippei~{\sc Fujisawa}\thanks{ifujisawa@particle.sci.hokudai.ac.jp} and
Ryuichi~{\sc Nakayama}\thanks{nakayama@particle.sci.hokudai.ac.jp}
       \\[1cm]
{\small
    Division of Physics, Graduate School of Science,} \\
{\small
           Hokkaido University, Sapporo 060-0810, Japan}
}
\date{
%\today \\
EPHOU-15-005  \\
March  2015
}
% 
%\begin{titlepage}
% 
\maketitle

\begin{abstract} 

Surface-charge algebra associated with BMS$_4$ symmetry on the null infinity of asymptotically flat spacetime is studied via the Hamiltonian framework. A coordinate system, where boundaries of constant-time hypersurfaces  cross the null infinity, is adopted. The equation itself which determines the variation of the surface charges turns out the same as that previously obtained via the covariant framework by Barnich and Troessaert, and is non-integrable for general radiation field $C_{AB}$. 
However, if $C_{AB}$ is independent of retarded time $u$, the variation equation is integrable and the conserved surface charges generate BMS$_4$ algebra without central extension.
\end{abstract}

%\end{titlepage}
\newpage
\setlength{\baselineskip}{18pt}

%%%%%%%%%%%%%%%%%%%%%%%%%%%%%%%%%%%%%%%%%%%%%%%%%%%%%%%%%%%%%%%%%%%%%
\newcommand{\bm}[1]{\mbox{\boldmath $#1$}}
\newcommand{\Slash}[1]{{\ooalign{\hfil/\hfil\crcr$#1$}}}
%%%%%%%%%%%%%%%%%%%%%%%%%%%%%%%%%%%%%%%%%%%%%%%%%%%%%%%%%%%%%%%%%%%%

\section{Introduction}
\hspace{5mm}
Minkowski spacetime has Poincar\'e group as an isometry. When the geometry fluctuates, then there is, in general, no exact isometry. It turns out, however, that when the metric tensor satisfies a certain appropriate asymptotic conditions, spacetime has an asymptotic symmetry.  At spatial infinity, there is asymptotic Poincar\'e symmetry. At null infinity the asymptotic symmetry is enhanced, and ordinary global translations are extended to local translation (supertranslation) in the retarded time $u$. The resulting asymptotic symmetry group,  which is composed of supertranslation and Lorentz group, is called BMS$_4$ group.\cite{BBM}\cite{Sachs}

In general relativity, conserved quantities such as Hamiltonian and angular momentum are represented as surface charges.\cite{Regge}\cite{BrownHenneaux}  
Energy and angular momentum of asymptotically flat spacetime at spatial infinity were represented as surface integrals.\cite{Komar}  
In 3 dimensional asymptotically anti-de Sitter spacetime, there is asymptotic infinite-dimensional conformal symmetry at the infinite boundary, and there exist associated surface charges  at the boundary. The Dirac bracket algebra of these charges (direct sum of two Virasoro algebras) was derived in the Hamiltonian framework.\cite{BrownHenneaux} 

On the null infinity of asymptotically flat spacetime, energy, momentum and angular momentum are not conserved due to non-vanishing fluxes via gravitational radiation.  Difficulty was recognized in defining surface charges associated with these non-conserved quantities on the null infinity.\cite{Wald} It was argued that Hamiltonian framework is not applicable for defining surface charges corresponding to  non-conserved quantities at null infinity, and that extra terms must be added to the variation equation for surface charges. By using covariant framework, such an extra term was proposed in \cite{Wald}. It was, however,  not addressed in \cite{Wald} whether the surface charges correctly generate the Dirac bracket algebra of BMS$_4$ symmetry. 

In \cite{BT2} Barnich and Troessaert carried out calculation of the Dirac bracket algebra of the extended BMS$_4$ by using covariant Lagrangian framework\cite{BB}. The extended BMS$_4$ group contains an infinite dimensional extension of the Lorentz group (superrotation), which may have singularities on the S$^2$ at null infinity.\cite{BT4}\cite{Barnich}\cite{Banks}\cite{Strominger3} In \cite{BT2} it was shown that the variation equation for the surface charges $Q$ is not integrable, and the Dirac bracket algebra of extended BMS$_4$\cite{BT2} at the null infinity was proposed by defining the algebra of the `integrable part' of the surface charges  by modifying the non-integrable variation $ \Slash{\delta} Q$. An interesting  proposal on algebra was made and it was shown that there appear central extensions which depend on the gravitational radiation field $C_{AB}$ on the null infinity. This proposal has not been justified. 

In the analysis via the covariant framework, however, the variation equation for the surface charges is still not integrable with the above modification of $ \Slash{\delta} Q$, and the following questions arise. 
Is it possible to obtain the  surface charges and their algebra in the Hamiltonian framework? 
Is the variation equation also non-integrable? 
The purpose of this paper is to show that the algebra of BMS$_4$ group at null infinity can be also studied via Hamiltonian framework, and to reconsider the algebra of surface charges.  
For this purpose we adopt a coordinate system, where  boundaries of  constant-time hypersurfaces cross the null infinity.  The surface charges are obtained as the boundary terms which must be added to the Hamiltonian in order to make the calculation of the Poisson bracket of Hamiltonians well-defined, even if partial integrations are required\cite{Regge}\cite{BrownHenneaux}. The variation of the surface charge, $\delta Q$, is uniquely determined and this defines the Dirac bracket algebra of the surface charges as $\{Q[\xi], Q[\xi']\}_D=\delta_{\xi'} Q[\xi]+K[\xi,\xi']$, where $K[\xi,\xi']$ is a central extension. 

In this paper, the constraint equation which determines the variation of the surface charges $\delta Q$ is obtained and found to be the same as the result \cite{BT2} in the covariant framework. This shows equivalence of the two frameworks for the present problem. It is confirmed that 
if a gravitational radiation field $C_{AB}$ depends on the retarded time $u$, surface charges are not conserved and the equation for $\delta Q$ is not  integrable. 
 While it is possible to add an appropriate additional term, {\em  i.e.}, (\ref{extra}) to the variation equation to cancel the non-integrable piece, it is not possible to make the charges generate appropriate symmetry algebra. If $C_{AB}$ does not depend on $u$, however, the variation equation becomes integrable. In this case the charges are conserved and generate symmetry algebra. 
Due to compatibility of the transformation rule of $C_{AB}$ and the  condition $\partial_u C_{AB}=0$, symmetry algebra is restricted to global BMS$_4$ and there is no central extension. On the other hand, if $\partial_u C_{AB} \neq 0$ and hence  charges are not conserved, appropriate charges which generate  BMS$_4$ algebra are not found.  Conclusion in this paper  is different from that in \cite{BT2}. 

We conclude this introduction with an outline of this paper. 
In sec.2 a coordinate system defined by foliation of spacelike hypersurfaces which are asymptotically Euclidean Anti-de Sitter \cite{SolodukindeBoer} is introduced. Each hypersurface has an infinite boundary on the null infinity. In sec.3 
Bondi frame metric without fixing the gauge of the metric component $g_{ut}$ is presented. The gauge of this component will not be fixed in this paper because except for gauge  $g_{ut}=(1/16r^2) C_{AB}C^{AB} +O(r^{-2})$ its fixing  generally breaks BMS$_4$ group. In sec.4 asymptotically flat metric in the coordinate system of sec.2 is obtained by a coordinate transformation. In sec.5 surface charge algebra and surface charges are obtained via Hamiltonian framework. It is shown that surface charges exist only for $C_{AB}$ independent of $u$. Sec.6 is left for summary.

\section{Asymptotically flat space via foliation of EAdS (and dS) hypersurfaces}
\hspace{5mm}
4d Minkowski space has two types of infinity. One is a spatial infinity (spi). 
In standard coordinates $ds_{\text{standard}}^2=-dt^2+dr^2+r^2 \, (d\theta^2+\sin^2 \theta d\phi^2)$  this is defined by a limit $r \rightarrow \infty$ with fixed time $t$. The other is a null infinity ${\cal I}$ \cite{Penrose1}. In coordinates, 
\begin{equation}
ds_0^2 =g_{\mu\nu} \, dx^{\mu} \, dx^{\nu}= -du^2-2du \, dr+r^2 \, (d\theta^2+\sin^2 \theta \, d\phi^2), \label{Minkowski1}
\end{equation}
where $u=t-r$ is a retarded time, it is defined by an unphysical metric obtained by a conformal rescaling with $\Omega=r^{-1}$.
\begin{equation}
d\tilde{s}_0^2=\tilde{g}_{\mu\nu} \, dx^{\mu} \, dx^{\nu}= \Omega^2 \, ds_0^2= -\Omega^2 \, du^2+2du \, d\Omega+ (d\theta^2+\sin^2 \theta \, d\phi^2) \label{ut}
\end{equation}
This metric $\tilde{g}_{\mu\nu}=\Omega^2 \, g_{\mu\nu}$ is regular and non-degenerate at infinity $\Omega=0$, and defines  differential and topological structures of ${\cal I}$. \footnote{This is a future null infinity, ${\cal I}^+$.There also exists a past null infinity, ${\cal I}^-$, defined with an advanced time $v=t+r$. In this paper we will concentrate on ${\cal I}^+$ and denote it as ${\cal I}$, since an extension to ${\cal I}^-$ is straightforward.}

Asymptotically flat spacetimes are defined in a similar way, and their unphysical metrics $\tilde{g}_{\mu\nu}$ are well defined and conformally flat at ${\cal I}$. These spacetimes are also characterized by an infinite-dimensional symmetry group.\cite{BBM}\cite{Sachs} While Minkowski space has Poincar\'e symmetry as an isometry,  curved spacetimes in general do not have isometries. However, asymptotically flat spacetimes have asymptotic symmetry group, BMS$_4$ group,  at ${\cal I}$. Penrose interpreted this group as conformal isometries of ${\cal I}$.\cite{Penrose2} BMS$_4$ group is a direct product of a supertranslation group and a Lorentz group.

Locally, Minkowski space can be represented as foliations of Euclidean Anti-de Sitter spaces (EAdS) and de Sitter spaces (dS). \cite{SolodukindeBoer} 
These foliations are defined (in the case of EAdS) by 
\begin{eqnarray}
ds_1^2 &=& -dX_0^2+dX_1^2+dX_2^2+dX_3^2, \nonumber \\
&& (-\tau^2=-X_0^2+X_1^2+X_2^2+X_3^2).
\end{eqnarray}
With a parametrization 
\begin{eqnarray}
X_0 &=& \tau \, \cosh \rho, \nonumber \\
X_1 &=& \tau \, \sinh \rho \, \sin \theta \cos \phi, \nonumber \\
X_2 &=& \tau \, \sinh \rho \, \sin \theta \sin \phi, \nonumber \\
X_3 &=& \tau \, \sinh \rho \, \cos \theta,  
\end{eqnarray}
we obtain the metric inside the future ($\tau >0$) and the past ($\tau <0$) regions of the lightcone put at some vertex (origin). 
\begin{equation}
ds_2^2 = -d\tau^2+\tau^2 \, (d\rho^2+\sinh^2 \rho \, (d\theta^2+\sin \theta \, d\phi^2))  \label{global}
\end{equation}
A hypersurface $\Sigma_{\tau}$ with constant $\tau$ is EAdS in global coordinates.

In the spacelike region outside the lightcone put at the origin we have a metric
\begin{equation}
ds_3^2 = d\sigma^2-\sigma^2 \, d\lambda^2+\sigma^2 \, \cosh^2 \lambda  \, (d\theta^2+\sin^2 \theta \, d\phi^2).
\end{equation}
This metric corresponds to a foliation in terms of dS hypersurfaces 
$-X_0^2+X_1^2+X_2^2+X_3^2=\sigma^2$ in global coordinates. 
The whole Minkowski spacetime is obtained by patching up these local coordinates constructed as foliations of EAdS$_3$ and dS$_3$. It is expected that (at least) some class of asymptotically flat spaces can also be obtained by foliations. In this paper this kind of asymptotically flat space  is studied. 

In coordinates $\{u,r,\theta,\varphi\}$, null infinity ${\cal I}^+$ is reached by taking limit $r \rightarrow +\infty$ with $u=t-r$ fixed. The retarded time $u$ ranges over $-\infty < u < \infty$. In coordinates $\{\tau, \rho, \theta, \varphi\}$, however, ${\cal I}^+$ is defined by $\rho \rightarrow \infty$ with $u=\tau \, e^{-\rho}$ fixed. In the forward lightcone time $\tau$ is positive, and only half $u>0$ of ${\cal I}^+$ is covered. Hence $u=0$ is a horizon in the local coordinates. The advantage of the coordinates (\ref{global}) over (\ref{ut}) is that constant-time hypersurfaces $\Sigma_{\tau}$ cross at infinite boundary ($\rho \rightarrow +\infty$) with ${\cal I}$. 
Hence the Hamiltonian framework can be used to study asymptotic symmetry algebra at ${\cal I}$.

\section{Symmetry of asymptotically-flat space in Bondi frame}
\hspace{5mm}
Asymptotically flat spacetime  in Bondi frame\cite{BBM}\footnote{The coordinate (\ref{B}) with $\beta=-(1/16) \, C^{AB}C_{AB}$ is called Bondi frame in the literature. Instead, hereafter we will call (\ref{B}) with any $\beta$ Bondi frame.} has the following  line element represented in the form of $1/r$ expansions. 
\begin{eqnarray}
ds_{\text{Bondi}}^2 &=& -( 1-\frac{2 m}{r}-\frac{2 m_1}{r^2} +\cdots) \, du^2 -2 \, (1+\frac{\beta}{r^2}+\cdots ) \, dudr \nonumber \\
&&-2(U_A +\frac{1}{r} \, W_A+\cdots) \, du \, dx^A+ \gamma_{AB} \, dx^Adx^B, \label{B}\\
\gamma_{AB} &=& r^2 \, h_{AB}+ r \, C_{AB}+H_{AB}+\cdots  \label{Bondi frame}
\end{eqnarray}
Here $A,B=z,\bar{z}$ and $x^z \equiv z$, $x^{\bar{z}} \equiv \bar{z}$ are complex coordinates for round $S^2$. It has standard metric, $h_{z\bar{z}}=2/(1+z\bar{z})^2$, $h_{zz} \equiv h_{\bar{z}\bar{z}}=0$. Fields 
$m$, $\beta$, $U_A$, $W_A$, $C_{AB}$, $H_{AB}$ are functions of $u$, $z$ and $\bar{z}$. 
$C_{AB}$ is traceless ($C_{z\bar{z}}=0$). These functions are determined by solving Einstein equation ($R_{\mu\nu}-\frac{1}{2} g_{\mu\nu}R=0$). 
\begin{eqnarray}
U_A&=&-\frac{1}{2} \, D^B \, C_{BA}, \label{U}\\
\dot{ m} &=& -\frac{1}{8} \, \dot{C}^{AB} \, \dot{C}_{AB}-\frac{1}{4} \, \partial_u^2 \, \Big({H^A}_A-\frac{1}{2} \, C^{AB} C_{AB}\Big) -\frac{1}{2}\, D_A \, \dot{U}^A, \label{dm} \\
\beta &=&-\frac{1}{4} \, {H^A}_A+\frac{1}{16} \, C^{AB}C_{AB}, \label{beta} \\
\frac{3}{2} \, \dot{W}_A &=& -D_A \, m+\frac{1}{2} \, D_A \dot{\beta}-\frac{1}{2} \, D^B\dot{H}_{BA}+\frac{1}{2} D_A{\dot{H}^C}\, _C-\frac{1}{2} \, D_B D^B U_A \nonumber \\
&& +\frac{1}{2} \, D_BD_A U^B+\frac{1}{2}\partial_u(C_{AB}U^B)-\frac{1}{4} \, \dot{C}^{BC}D_AC_{BC} \nonumber \\ &&-\frac{1}{2} C^{BC}D_A \dot{C}_{BC}+\frac{1}{2} \, D^B \, \big(C_{BC} \, {{\dot{C}^C\,}}_A \big), \label{W}\\
H_{zz} &=& H_{zz}(z), \quad H_{\bar{z}\bar{z}}=H_{\bar{z}\bar{z}}(\bar{z}). \label{Hzz}
\end{eqnarray}
Here dot means a derivative with respect to $u$. Indices $A$, $B$... are raised and lowered by means of $h_{AB}$ and its inverse  $h^{AB}$. $D_A$ is a covariant derivative with respect to $h_{AB}$, and non-zero Christoffel symbols are given by $\Gamma^z_{zz}=-2\, \bar{z}\, (1+z\bar{z})^{-1}$ and $\Gamma^{\bar{z}}_{\bar{z}\bar{z}}=-2z\, (1+z\bar{z})^{-1}$. There is also a $u$-flow equation for $m_1$, which is not presented here. 

Function $\beta$ is gauge-fixed in several ways in the literature.\footnote{Our $\beta$ is different from that in \cite{BBM}.} For instance, a gauge $\beta=0$ is chosen in the null tetrad formalism\cite{Newman}. Another gauge $\beta= -(1/16) \, C_{AB}C^{AB} $ is chosen in \cite{BBM} so that $\sqrt{\text{det} \gamma_{AB}}=r^2 \, \sqrt{\text{det}h_{AB}}$.  It can be shown that gauge condition such as 
\begin{equation}
\beta=\lambda \, C^{AB} \, C_{AB},   \label{beta gauge}
\end{equation}
where $\lambda$ is a constant, cannot be preserved under BMS$_4$ group except for a value $\lambda=-1/16$. That is,  gauge fixing  (\ref{beta gauge}) generally breaks BMS$_4$ group. This will be shown at the end of this section. For this reason $\beta$ will {\em not} be fixed in this paper. 
The eqs of motion (\ref{U})-(\ref{Hzz}) for gauge $\beta=-\frac{1}{16} \, C_{AB}C^{AB}$ are obtained in \cite{BBM}\cite{Sachs}\cite{Barnich}.

Asymptotically flat spacetime has BMS$_4$ group as asymptotic symmetry. This is composed of a group of (i) supertranslation and that of  (ii) superrotation. Supertranslation (i) is a group of local translation in retarded time $u$. The corresponding asymptotic Killing vector $\xi^{\mu}$ is given by \cite{Sachs}\cite{BT2} 
\begin{eqnarray}
\delta u &= & \xi^u \equiv f,   \\
\delta r &=& \xi^r\equiv D^zD_z f +\frac{1}{r} \, (U^A \, D_A f-\frac{1}{4} \, C^{AB} \, D_A D_B f)+\cdots, \\
\delta z &=& \xi^z\equiv -\frac{1}{r} \, D^z f+\frac{1}{2 r^2} \, C^{zz} \, D_z \, f \label{z1}+\cdots, \\
\delta \bar{z} &=&\xi^{\bar{z}}\equiv -\frac{1}{r} \, D^{\bar {z}} f+\frac{1}{2 r^2} \, C^{\bar{z}\bar{z}} \, D_{\bar{z}} \, f+\cdots.  \label{zb1}
\end{eqnarray}
Here $f=f(z,\bar{z})$ is a scalar function on a sphere. Subleading terms ($\cdots$) are suppressed here.

Superrotation (ii) corresponds to extension of Lorentz transformation in Minkowski space and has an asymptotic Killing vector, \cite{Sachs}
\begin{eqnarray}
\delta u &=& \xi^u \equiv \frac{1}{2} \, u \, D_z \zeta^z, \\
\delta r &=& \xi^r \equiv -\frac{1}{2} \,  (r+u) \, D_z \zeta^z+\frac{u}{2r} \, \Big(U^z \, D_z^2 \zeta^z-U_z \, \zeta^z-\frac{1}{4} \, C^{zz} \, D_z^3\zeta^z\Big)+\cdots, \\
\delta z &=& \xi^z \equiv \zeta^z+\frac{u}{2r}\, \zeta^z -\frac{u}{4r^2} \, C^{zz} \, D^2_{z} \, \zeta^{z}+\cdots, \label{z2} \\
\delta \bar{z} &=& \xi^{\bar{z}} \equiv -\frac{u}{2r} \, h^{z\bar{z}} \, D^2_z \, \zeta^{z} -\frac{u}{4r^2} \, h_{z\bar{z}}\, C^{\bar{z}\bar{z}} \,\zeta^{z}+\cdots. \label{zb2}
\end{eqnarray}
Here $\zeta^z=\zeta^z(z)$ is a polynomial of $z$ up to quadratic order.  There is also a similar transformation with a parameter function $\zeta^{\bar{z}}=\zeta^{\bar{z}}(\bar{z})$.  For $\zeta^z=1,i,z,iz, z^2,iz^2$, 
both  transformations generate $SL(2,C)$ group. 
It is argued that the transformation functions $\zeta^z(z)$ and $\zeta^{\bar{z}}(\bar{z})$ may be generalized to arbitrary (anti-) holomorphic functions, albeit singularities on $S^2$.\cite{Barnich}\cite{Strominger3} Then one obtains a direct product of two Virasoro groups. In this paper we will consider this extended symmetry group, and call it simply BMS$_4$ group. 
Starting from (\ref{Bondi frame}), these transformations preserve the following asymptotic flatness condition.
\begin{eqnarray}
g_{uu} & \sim & -1+ {\cal O}(r^{-1}), \label{ASF1}\\
g_{ur} & \sim & -1+ {\cal O}(r^{-2}), \label{ASF2}\\
g_{AB} & \sim & r^2 \, h_{AB}+  {\cal O}(r^{1}), \label{ASF3}\\
g_{uA} & \sim &  {\cal O}(1), \\
g_{rA} &= &  0, \\
g_{rr} &= &  0 \label{ASF6}
\end{eqnarray}

The function $\beta$ is the subleading term of (\ref{ASF2}): $g_{ur}=-1-\frac{\beta}{r^2}+\cdots$. Under BMS$_4$ group, $g_{ur}$ transforms as $\delta_{\xi} \, g_{ur}=\nabla_u \, \xi_r+\nabla_r \, \xi_u$. This gives a transformation rule of $\beta$. In the case of supertranslation, 
\begin{equation}
\delta_f \, \beta= f \, \partial_u \, \beta+\frac{1}{4} \, C^{AB} \, D_AD_B \, f.
\label{delfBeta}
\end{equation}
For superrotation, we have 
\begin{equation}
\delta_{\zeta} \, \beta= \zeta^z \, \partial_z \, \beta+D_z \zeta^z \, \beta+\frac{1}{2} \, D_z\zeta^z \, u\partial_u \, \beta+\frac{1}{8} \, u \, C^{zz} \, D_z^3\, \zeta^z.  \label{delzetaBeta}
\end{equation}
Only for $\lambda=-\frac{1}{16}$, the gauge fixing condition (\ref{beta gauge}) is compatible with (\ref{delfBeta}) and (\ref{delzetaBeta}) and the transformation rules of $C_{AB}$, with are also derived via a definition $g_{zz} \equiv r \, C_{zz}+O(r^0)$ and $\delta g_{zz}=2\nabla_z \, \xi_z$,
\begin{eqnarray}
\delta_f \, C_{AB} &=& f \, \partial_u \, C_{AB}-2D_AD_B \,  f  +h_{AB} \, D_CD^C f, \label{delfC}\\
\delta_{\zeta} \, C_{zz} &=& \zeta^z \, D_z \, C_{zz}+2D_z \zeta^z \, C_{zz}+\frac{1}{2} \, D_z\zeta^z \, (u\partial_u-1) \, C_{zz}-u D_z^3 \zeta^z, \label{delzCzz} \\
\delta_{\zeta} \, C_{\bar{z}\bar{z}} &=& \zeta^z D_zC_{\bar{z}\bar{z}}+\frac{1}{2} \, D_z\zeta^z \, (u\partial_u-1) \, C_{\bar{z}\bar{z}}. \label{delzCzbzb}
\end{eqnarray}
Otherwise,  gauge fixing (\ref{beta gauge}) with $\lambda \neq -1/16$ cannot be used so that $\beta$ must be left unfixed. 

\section{Asymptotic symmetry in new coordinate system}
\hspace{5mm}
An asymptotically flat spacetime, which is described by coordinates $\{\tau,\rho,z,\bar{z}\}$ which are deformation of local coordinates  (\ref{global}) of Minkowski space, also has asymptotic symmetry. 
A coordinate transformation which connects the two metrics in both coordinates $\{r,u,z,\bar{z} \}$ and $\{\tau,\rho,z,\bar{z}\}$  is given by 
\begin{eqnarray}
u &=& \tau \, e^{-\rho}, \label{tr1}\\
r &=& \tau \, \sinh \rho. \label{tr2}
\end{eqnarray}
The line element in coordinates $\{\tau, \rho, z,\bar{z}\}$ is obtained by substituting (\ref{tr1})-(\ref{tr2}) into (\ref{Bondi frame}). We will use a line element obtained from (\ref{Bondi frame}) by this transformation, instead of solving Einstein equation from scratch. Whether this new line element has BMS$_4$ symmetry or not must be checked once again independently, since the coordinate transformation changes the orders of terms in the variation of the line element, and does not enssure it. This will be discussed soon. 

Half of  future null infinity ${\cal I}^+$ ($u>0)$ is reached in the limit $r \rightarrow +\infty$ with $u$ fixed. This requires a fine-tuned limit in the $(\tau,\rho)$ coordinates.
\begin{equation}
\tau, \ \rho \rightarrow +\infty, \qquad 
\tau \, e^{-\rho} =\, \text{fixed} \, (=u)   \label{doublescaling}
\end{equation}
In this way the line element (\ref{Bondi frame}) will be rearranged into $1/r \sim e^{-2\rho}$ expansions, where coefficients are functions of $u$. 

Line element in this frame obtained by the above coordinate transformation is given by 
\begin{equation}
ds^2 = G_{\mu\nu} \, dx^{\mu}dx^{\nu},
\end{equation}
where $\mu, \nu=\tau, \rho, z,\bar{z}$ and 
\begin{eqnarray}
G_{\tau\tau} &=& -1+\frac{4m}{\tau}\frac{e^{-3\rho}}{1-e^{-2\rho}}-\frac{4\beta}{\tau^2}\frac{e^{-2\rho}}{1-e^{-2\rho}} 
+\cdots, \label{MetricG1}\\
G_{\tau \rho} &=& -\frac{4m}{1-e^{-2\rho}} e^{-3\rho}-\frac{4\beta e^{-4\rho}}{\tau(1-e^{-2\rho})^2}
+\cdots, \\
G_{\tau A} &=& 
-e^{-\rho} \, \big[U_A+  \frac{2e^{-\rho}}{\tau \, (1-e^{-2\rho} )}\, W_A\big]+\cdots, \\
G_{\rho A} &=& \tau \,e^{-\rho} \, \big[U_A+\frac{2e^{-\rho}}{\tau \, (1-e^{-2\rho} )}\, W_A \big]+\cdots, \\
G_{\rho \rho} &=& \tau^2+\frac{4m\tau}{1-e^{-2\rho}} e^{-3\rho}+\frac{4\beta e^{-2\rho}}{(1-e^{-2\rho})^2}(1+e^{-2\rho}) 
+\cdots, \\
G_{AB} &=& \frac{1}{4} \, \tau^2 \, e^{2\rho}(1-e^{-2\rho})^2 \, h_{AB}+
            \frac{1}{2} \tau \, e^{\rho} (1-e^{-2\rho}) \, C_{AB}+H_{AB}+\cdots \label{MetricG2}
\end{eqnarray}
Fields, $C_{AB}$, $U_A$, $W_A$, $\beta$  and $m$, have coordinate dependence like $C_{AB}(\tau \, e^{-\rho}, z,\bar{z})$. 
Since the above metric (\ref{MetricG1})-(\ref{MetricG2}) is obtained from Bondi-frame metric by coordinate transformation, it also satisfies Einstein equation.  
If $u=\tau \, e^{-\rho} (> 0)$ is kept fixed near large-$\rho$ region, the above metric takes the form of $1/e^{2\rho}$ expansion. Coefficient functions in the expansion of the metric tensor contain terms with negative powers of $u$. 
Later, we will use this $1/e^{2\rho}$ expansion to derive a variation formula of surface charges $\delta \, Q$ at $\rho \rightarrow \infty$.
It is checked that when $\tau (\neq 0)$ is kept finite, terms which diverge as $\rho \rightarrow \infty$ cancel out. Since $u^{-1}= \tau^{-1} \, e^{\rho}$, this means that  negative powers of $u$ do not appear in $\delta Q$.  On the other hand, terms proportional to $u=\tau e^{-\rho}$ and $u^2=\tau^2 e^{-2\rho}$  may be  created, by keeping terms which are higher orders in $e^{-\rho}$.

Under supertranslation, $\tau$ and $\rho$ change as 
\begin{eqnarray}
\delta \, \tau &=& f \, \cosh \rho+e^{-\rho} \, D^zD_z f+ O(\tau^{-1} e^{-2\rho}),  \label{taurho1}\\
\delta \, \rho &=& -f\tau^{-1} \sinh \rho+\tau^{-1}e^{-\rho} D^zD_z f + O(\tau^{-2} e^{-2\rho}). \label{taurho2}
\end{eqnarray}
Since $\delta \tau$ is large for finite $\tau$ and extremely large $\rho$, supertranslation is not a symmetry at ${\cal I}$ for finite $\tau$. However, when $\tau$ is taken to infinity at the same time, it becomes an asymptotic isometry. 
The line element for Minkowski space in the new frame (\ref{global}) changes as follows.
\begin{equation}
\delta \, (ds_2^2) = -2\tau \, \sinh \rho \, D_z^2 f\, dz^2+2 \, e^{-\rho}(D_zf+D_z^2D^zf) \, dz \, (\tau \, d\rho-d\tau)+ \text{c.c} \label{delS0}
\end{equation}
In the case of Bondi frame there is only a single term, $-2r D_z^2f dz^2$, corresponding to the first term on the right above. Here an extra term (the second term) appears. These changes can be compensated by the changes of $C_{AB}$ and $U_{A}$. The above change (\ref{delS0})  vanishes for the four global translations $f=1, z/(1+z\bar{z}), \bar{z}/(1+z\bar{z}), (1-z\bar{z})/(1+z\bar{z})$, as it should. 

Under superrotation, transformations of $\tau$ and $\rho$ are simply, 
\begin{eqnarray}
\delta \, \tau &=&  O(e^{-3\rho}), \label{taurho3}\\
\delta \, \rho &=& -\frac{1}{2} \, D_A \, \zeta^A+ O(\tau^{-1} e^{-3\rho}).  \label{taurho4}
\end{eqnarray}
Hence superrotation is an asymptotic symmetry also in this frame. 
In this case the change of the line element is 
\begin{equation}
\delta \, (ds^2_2)= -\tau^2 \, e^{-\rho} \, \sinh \rho \, D_z^3\zeta^z \, dz^2+\text{c.c}.
\end{equation}

\section{Surface charge algebra of BMS$_4$}
\hspace{5mm}
The surface charges (generators) of BMS$_4$ group in four dimensions and their algebra have  previously been studied via covariant approach in \cite{Barnich}\cite{BT2}. In this section, this task will be carried out by using the Hamiltonian framework\cite{Regge}\cite{BrownHenneaux}\cite{Kubota}. This is based on Hamiltonian (ADM) formulation of gravity\cite{ADM}. We  choose $\tau$ as a time variable and a foliation of the spacetime with $\tau >0$ into constant-time space-like slices $\Sigma_{\tau}$.  $\Sigma_{\tau}$ has a time-like normal vector $n_{\mu}=-\delta^{\tau}_{\mu}+{\cal O}(e^{-2\rho})$. 
Each slice $\Sigma_{\tau}$ is a 3D Euclidean Anti-de Sitter space (EAdS$_3$) and has boundary at $\rho=+\infty$. $\rho=0$ is the center of this space. When taking the asymptotic limit $\rho \rightarrow +\infty$, we need to introduce a large cutoff $\rho=\rho_{\infty}$ in order to regularize divergences, and we obtain a cylinder with a radius $\rho_{\infty}$ and a height in the direction of increasing $\tau$. The time-like boundary at $\rho=\rho_{\infty}$ will be denoted as $\Sigma_{\rho_{\infty}}$. 

In order to reach a point with positive value of $u$ on ${\cal I}^+$, $\tau$ must also be increased according to (\ref{doublescaling}), as the cutoff is removed, $\rho_{\infty} \rightarrow +\infty$. In this case one is forced to approach ${\cal I}^+$ along a null geodesic $u=$ const.\footnote{If we do not shift to hypersurfaces $\Sigma_{\tau}$ with increasing $\tau$ appropriately, then we will end up with the point $u=0$ of ${\cal I}^+$.} This, however, does not imply that canonical formalism on a null hypersurface is considered. Canonical commutation relations are imposed on a spacelike hypersurface $\Sigma_{\tau}$, and $\tau, \rho \rightarrow \infty$ limit will be taken afterwords.  

\subsection{Hamiltonian}
\hspace{.5cm}
In ADM decomposition\cite{ADM}, the spacetime metric is arranged into the form
\begin{equation}
ds^2= G_{\mu\nu} \, dx^{\mu} \, dx^{\nu}=-(N^2-N_a N^a) \, d\tau^2
+\gamma_{ab} \, dx^a \, dx^b+2N_a \, dx^a \,d\tau, \label{ADM}
\end{equation}
where $a$ runs over $\rho, z,\bar{z}$, and $N^a=\gamma^{ab} \, N_b$ with $\gamma^{ab}$ being the inverse of $\gamma_{ab}=G_{ab}$. 
In the case of the present solution, the lapse and shift functions $N$, $N_a$ depend on the fields.   
\begin{eqnarray}
N_{\rho} &=& -4m e^{-3\rho}-(4m e^{-5\rho}+\frac{4\beta}{\tau} e^{-4\rho})+\cdots,  \\
N_{A} &=& -e^{-\rho}U_A-\frac{2}{\tau} W_A (e^{-2\rho}+e^{-4\rho})+\cdots, \\
N &=& 1+(-\frac{2m}{\tau} e^{-3\rho}+\frac{2\beta}{\tau^2} e^{-2\rho})+\cdots \label{shiftA}
\end{eqnarray}

Action integral $S$ is given by 
\begin{equation}
S= \frac{1}{16\pi G} \, \int d^4x \, \sqrt{-G} \, R. 
\end{equation}
The momentum conjugate to $\gamma_{ab}$ is given by 
\begin{equation}
\Pi^{ab}= \frac{1}{16\pi G} \, \sqrt{\gamma} \, (K^{ab}-\gamma^{ab} \, K),
\end{equation}
where $K^{ab}$ is extrinsic curvature of $\Sigma_{\tau}$: $K_{ab}= (1/2N)(\partial_{\tau} \gamma_{ab}-N_{a|b}-N_{b|a})$. 
Symbol \lq $|a$' stands for a covariant derivative with respect to $\gamma_{ab}$. 
On a spacelike surface $\Sigma_{\tau}$, canonical commutation relations are imposed on $\gamma_{ab}$ and $\Pi^{ab}$. 
The Hamiltonian is given by  
\begin{eqnarray}
H[N,N^a]&=& \int_{\Sigma_{\tau}} d\rho d^2z \, \Big[16\, \pi \, G\gamma^{-1/2} \, N  \,  \big(\Pi^{ab}\Pi_{ab}-\frac{1}{2} \, \Pi^2\big)- \frac{1}{16\pi G} \, \sqrt{\gamma} \, N ~^{(3)}R\Big] \nonumber \\
&& -  \int_{\Sigma_{\tau}} d\rho d^2z \sqrt{\gamma} \, 2\, N_a \, \Big(\frac{1}{\sqrt{\gamma}}\, {\Pi^{ab}}\Big)_{|b}.
\end{eqnarray}
$^{(3)}R$ is the Ricci scalar constructed from $\gamma_{ab}$, and  $\Pi= \gamma_{ab} \, \Pi^{ab}$.

The generator ${\cal H}[\xi]$ of the asymptotic symmetry is obtained by replacing $N$ and $N^a$ in $H[N,N^a]$, by $\bar{\xi}^{\tau}$ and 
$\bar{\xi}^a$ as 
\begin{eqnarray}
N & \rightarrow & \bar{\xi}^{\tau} \equiv N \, \xi^{\tau}, \nonumber \\
N^a & \rightarrow & \bar{\xi}^a \equiv N^a \, \xi^{\tau}+ \xi^a,
\end{eqnarray}
where $\xi^{\mu}=(\xi^{\tau}, \xi^a)$ are equal to $(\delta \tau, \delta \rho, \delta z, \delta \bar{z})$ in (\ref{z1})-(\ref{zb1}), (\ref{z2})-(\ref{zb2}) and (\ref{taurho1})-(\ref{taurho4}) which correspond to supertranslation $\xi^{\mu}=(\delta r, \delta u, \delta z, \delta \bar{z})$, and superrotation re-expressed in $\{\tau,\rho, z,\bar{z}\}$ coordinates. 

\subsection{Variation equation for surface charges}
\hspace{.5cm}
Transformations of the canonical variables are expressed in terms of Poisson brackets: $\delta_{\xi} \gamma_{ab}= \{ \gamma_{ab}, {\cal H}[\xi]\}$, $\delta_{\xi} \, \Pi^{ab}= \{ \Pi^{ab}, {\cal H}[\xi]\}$.  To compute the Poisson brackets, partial integration is in general necessary, and to make the Poisson bracket well-defined, an appropriate surface term $Q[\xi]$ must be added to ${\cal H}[\xi]$. Its variation must satisfy the following variation constraint.
\begin{multline}
\delta \, Q[\xi] =
\int_{S} d^2z \, \Big[ 2\, \bar{\xi}^a \, \Pi^{b\rho} \, \delta \gamma_{ab}+2\bar{\xi}^a \, \delta \, \Pi^{b\rho}  \, \gamma_{ab} -\bar{\xi}^{\rho} \, \Pi^{ab} \, \delta \, \gamma_{ab}\Big] \\
+\frac{1}{16\pi G} \, \int_S d^2z \, \sqrt{\gamma} \, S^{abc\rho} \, \Big[\ \bar{\xi}^{\tau} \, (\delta \, \gamma_{ab})_{|c}-\partial_c \, \bar{\xi}^{\tau} \, \delta \, \gamma_{ab}\Big]  \label{delQ}
\end{multline}
Here $S$ is 2-dim sphere at fixed $u$ and $r \rightarrow \infty$. 
$S^{abcd}$ is defined by 
\begin{equation}
S^{abcd}= \frac{1}{2} \, (\gamma^{ac} \, \gamma^{bd}+\gamma^{ad} \, \gamma^{bc}-2 \, \gamma^{ab} \, \gamma^{cd}).
\end{equation}

Hamilton's eqs  are given as follows.
\begin{eqnarray}
\delta_{\xi} \, \gamma_{ab} &=& \bar{\xi}_{a|b}+\bar{\xi}_{b|a}+\frac{32\pi G}{\sqrt{\gamma}} \, \bar{\xi}^{\tau}\, (\Pi_{ab}-\frac{1}{2} \, \gamma_{ab} \, \Pi), \\
\delta_{\xi} \, \Pi^{ab} &=& -\frac{32\pi G}{\sqrt{\gamma}} \, \bar{\xi}^{\tau} \, \big(\Pi^{ac} \, {\Pi_c}^b-\frac{1}{2} \, \Pi \, \Pi^{ab}\big)
+\frac{8\pi G}{\sqrt{\gamma}} \, \bar{\xi}^{\tau}\big(\Pi^{cd}\Pi_{cd}-\frac{1}{2}\Pi^2\big) \, \gamma^{ab} \nonumber \\
&&-\frac{1}{16\pi G} \, \sqrt{\gamma} \, \bar{\xi}^{\tau} \, 
\big(^{(3)} R^{ab}-\frac{1}{2}\gamma^{ab} \, ^{(3)} R\big)+\frac{1}{16\pi G}\sqrt{\gamma}\big(\bar{\xi}^{\tau |ab}-\gamma^{ab}{\bar{\xi}^{\tau |c}} \, _c\big)
\nonumber \\
&& +\sqrt{\gamma} \, \Big(\frac{1}{\sqrt{\gamma}} \, \Pi^{ab} \, \bar{\xi}^c \Big)_{|c}-\Pi^{ac} \, {N^b}_{|c}-\Pi^{bc} \, {N^a}_{|c}
\end{eqnarray}

In the case of AdS$_3$, the Hamiltonian generator ${\cal H}[\xi]$ generates a Poisson bracket algebra:
\begin{equation}
\ \{{\cal H}[\xi], {\cal H}[\xi']\} ={\cal H}[[\xi,\xi']]+K[\xi,\xi'],
\end{equation}
where $K[\xi,\xi']$ is a possible central extension. A strategy to evaluate this central extension is  to replace Poisson bracket by Dirac bracket. Then the Hamiltonian and momentum constraints hold strongly, and the generator ${\cal H}[\xi]$ can be replaced by the surface charge $Q[\xi]$:
\begin{equation}
\ \{Q[\xi], Q[\xi']\}_D =Q[[\xi,\xi']]+K[\xi,\xi'].
\end{equation}
The left hand side of this equation may be evaluated as $\delta_{\xi'} \, Q[\xi]$, where $\delta_{\xi'}$ is a transformation associated with an asymptotic Killing vector $\xi'$. For AdS background, $Q[[\xi,\xi']]$ on the right hand side vanishes. Hence the central charge is evaluated as 
\begin{equation}
K[\xi,\xi'] = \delta_{\xi'} \, Q[\xi]\big|_{\text{AdS}}.  \label{defK}
\end{equation}
Here $\delta_{\xi'}$ stands for a variation associated with transformation $\xi'$.

\subsection{Charge algebra}
\hspace{.5cm}
Same prescription is  used in the case of a four-dimensional asymptotically flat space on ${\cal I}$. The surface charges are denoted as $Q[f,\zeta^z, \bar{\zeta}^{\bar{z}}]$, where $f(z,\bar{z})$ is a parameter function for supertranslation and $\zeta^z(z)$, $\bar{\zeta}^{\bar{z}}(\bar{z})$ those for (extended) superrotations. The brackets of surface charges are given by 
\begin{equation}
\ \Big\{Q[\xi], Q[\xi']\Big\}_D = \delta_{\xi'} \, Q[\xi]  \label{QQdelQ}
\end{equation}
for $\xi=(f,\zeta, \bar{\zeta})$ and $\xi'=(f',\zeta', \bar{\zeta}')$. 
Calculation is done by using Mathematica and only the results are now presented.\footnote{Here (non-)integrability of eq (\ref{delQ}) is not yet taken into account.}\footnote{In eq (\ref{tr}), the last equality is checked by actual calculation.}
\begin{multline}
\ \Big\{ Q[f_1,0,0], Q[f_2,0,0] \Big\}_D \equiv  \delta_{f_2} \, Q[f_1,0,0]  \\
= \frac{1}{16\pi G} \, \int_S d^2z \, h_{z\bar{z}} \, 
\dot{C}^{AB}  \, (f_2 \,D_AD_B \, f_1-f_1 \, D_AD_B \, f_2),
\label{tt}
\end{multline}
\begin{multline}
\ \Big\{ Q[0,\eta,0], Q[f,0, 0] \Big\}_D \equiv  \delta_{f} \, Q[0,\eta,0]  \\
= \frac{1}{16\pi G} \, \int_S d^2z \, h_{z\bar{z}} \, [\eta,f] \, \Big[4 \, m+\partial_u \, {H_A}^A -\frac{1}{2} \partial_u \, (C_{AB}  \,
C^{AB}) \Big] \\
+\frac{1}{16\pi G} \, \int_S d^2z \, h_{z\bar{z}} \, \frac{1}{2} \Big[f \, D_z^3 \, \eta^z \, (u\partial_u-1)C^{zz}- u \, D_z\eta^z \, D_AD_B \, f \, \dot{C}^{AB}\Big]  \\
+\frac{1}{16\pi G} \, \int_S d^2z \, h_{z\bar{z}} \, \Big[-\frac{3}{4}fD_z\eta^z \,  \dot{ C}^{zz} \, C_{zz} -\frac{1}{2} \, f\eta^z \dot{C}^{zz} \, D_z C_{zz}\\+\frac{1}{4}fD_z\eta^z  \,  \dot{C}_{zz} \, C^{zz}-\frac{1}{2} \, f  \eta^z \, \dot{C}_{zz} D_zC^{zz}\Big], \label{rt}
\end{multline}
\begin{equation}
\ \Big\{ Q[f,0,0], Q[0, \chi,0] \Big\}_D \equiv  \delta_{\chi} \, Q[f,0,0]  =-\delta_f \, Q[0,\chi,0], \label{tr}
\end{equation}
\begin{multline}
\ \Big\{Q[0,\eta,0], Q[0,\chi,0]\Big\}_D \equiv \delta_{\chi}  \, Q[0,\eta,0]  \\
= \frac{1}{16\pi G} \, \int_S d^2 z h_{z\bar{z}} \, [\eta,\chi]^z \, \Big[ -3 W_z+C_{zz} \, U^z+\frac{3}{4} \, D_z \, {H_A}^A-\frac{5}{16} \, D_z(C_{AB} C^{AB})  \\
-2u \, \partial_z \, m -\frac{1}{2} \, u \, D_z \, {\dot{H}_A}^A \\
+\frac{1}{4} \, u \,  (2\, C^{zz} \, D_z \, \dot{C}_{zz} +\dot{C}_{zz} \, D_z \, C^{zz}+2\, C_{zz} \, D_z \, \dot{C}^{zz} +\dot{C}^{zz} \, D_z C_{zz}) \Big]\\
-\frac{1}{64\pi G} \, \int_S d^2 z h_{z\bar{z}} \, (D_z\eta^z \, D_z^3\chi^z-D_z\chi^z \, D_z^3 \, \eta^z) \, u \,  (u\partial_u-1) \, C^{zz}
, \label{rr} 
\end{multline}
\begin{multline}
\ \Big\{ Q[0,0,\bar{\chi}], Q[0,\eta,0]\Big\}_D \equiv \delta_{\eta} \, Q[0,0,\bar{\chi}]  \\
=\frac{1}{16 \pi G} \, \int_S d^2z h_{z\bar{z}} \, \Big[\frac{1}{4} u\, D_z^3\eta^z \, D_{\bar{z}} \chi^{\bar{z}} \, (1-u\partial_u) \, C^{zz}-\frac{1}{4} u\, 
D_{\bar{z}}^3\chi^{\bar{z}} \, D_{z} \eta^z\, (1-u\partial_u) \, C^{\bar{z}\bar{z}}\\
+\frac{1}{2}u \, D_z\eta^z D_{\bar{z}}\chi^{\bar{z}} \, (C_{zz}\, \dot{C}^{zz}-\dot{C}_{zz}\, C^{zz})
+\frac{1}{4} u\, \eta^z \, D_{\bar{z}} \chi^{\bar{z}}(\dot{C}_{zz}\, D_zC^{zz}+ \dot{C}^{zz}\, D_zC_{zz}) \\
-  \frac{1}{4} u\, \chi^{\bar{z}} \, D_{z} \eta^{z}(\dot{C}_{zz}\, D_{\bar{z}}C^{zz}+ \dot{C}^{zz}\, D_{\bar{z}}C_{zz})            \Big].
\label{rbr}
\end{multline}
Here, $[\eta,f] \equiv -[f,\eta] \equiv \eta^z \, \partial_z \, f-\frac{1}{2} \, f \, D_z \eta^z$ and $[\eta,\chi]^z \equiv \eta^z \, \partial_z \, \chi^z-\chi^z \, \partial_z \, \eta^z$. There are also similar brackets involving $Q[0,0,\bar{\chi}]$ instead of $Q[0,\eta,0]$, which are not displayed here.

In the above calculation, there appeared terms proportional to $e^{\rho} \, \tau$, which are divergent as $\rho \rightarrow +\infty$.  These can be dropped after partial integration over S$^2$, and the charges are finite.  There also exist terms proportional to $\tau^{-2}$. However, those terms vanish in the $\tau \rightarrow +\infty$ limit. 
In the calculation of $\{Q[0,\eta,0], Q[0,\chi,0]\}_D$ there is also a term 
proportional to $h_{z\bar{z}} \, \chi^z(z) \, \eta^z(z) \, H_{zz}(z)$ in the integrand. 
This is formally dropped by using a formula $D_{\bar{z}} \, D_z \, \eta^z(z)=-h_{z\bar{z}} \, \eta^z(z)$ and a partial integration with respect to the derivative $D_{\bar{z}}$. \footnote{Rigorously speaking, we must be careful in this procedure, because the surface terms for $z, \bar{z}$ integration cannot be dropped for polynomials $\eta^z$ and $\chi^z$. This issue needs to be further studied. This term might as well be dropped by assuming $H_{zz}(z)=0$.   }

The above result shows that the variation of surface charge $\delta_{\xi} \, Q[\xi']$ satisfies a condition of anti-symmetry.
\begin{equation}
\delta_{\xi} \, Q[\xi'] =-\delta_{\xi'} \, Q[\xi] \label{antisym}
\end{equation}
This is a necessary condition for prescription (\ref{QQdelQ}) to work.

\subsection{Non-integrability of variation equation for surface charges}
\hspace{5mm}
Surface charges are defined by variation equation (\ref{delQ}). Integrability of this equation must be checked. 
For arbitrary variations $\delta$ of fields, eq (\ref{delQ}) for supertranslation reads
\begin{equation}
\delta Q[f,0,0] = \frac{1}{16\pi G} \, \int_S d^2z \, h_{z\bar{z}} \, f \, \big[4 \delta M+\frac{1}{2} \, \dot{C}_{AB} \, \delta \, C^{AB}\big],  \label{delQf}
\end{equation}
and that for superrotation reads\footnote{On the right hand side of (\ref{delQeta}) there also exists  a divergent term in the integrand:\\ $-e^{2\rho} \,u \, h_{z\bar{z}} \, \eta^z(z) \, \delta \, U_z$. This term, however,  drops out after partial integration. }
\begin{multline}
\delta Q[0,\eta,0] =\frac{1}{16\pi G} \, \int_S d^2z \, h_{z\bar{z}} \, \eta^z \, \Big[\delta (-2L_z-2uD_z M)-\frac{1}{4} \, u \, D_z(\dot{C}_{AB} \, \delta \, C^{AB})\Big].   \label{delQeta}
\end{multline}
These equations for variations of charges coincide with those obtained via covariant framework\cite{BT2}.
Here $M$ is a gauge invariant mass,\footnote{Definition $\delta_f \, g_{uu} \rightarrow 2\delta_fm/r$ yields $\delta_f m=f\dot{m}-\dot{U}^AD_Af+(1/4) \dot{C}^{AB}D_AD_B f$. This is true only when $\beta=-(1/16)C_{AB}C^{AB}$. For other choice of $\beta$, it is $M$, not $m$, that transforms as (\ref{delfM}). Similar statement is valid for $\delta_{\zeta} m$ and $\delta_{\zeta} \, M$.}
 {\it i.e.}, mass which does not depend on the choice of $\beta$,
\begin{equation}
M \equiv m+\frac{1}{4} \partial_u {H_A}^A-\frac{1}{8} \, \partial_u (C_{AB}C^{AB}).
\end{equation}
This transforms under supertranslation as 
\begin{eqnarray}
\delta_f \, M &=& -\frac{1}{8} \, f \, \partial_u C_{AB} \partial_u C^{AB}+\frac{1}{4} \, D^AD^B (f \partial_u C_{AB}) \nonumber \\
&=& f \, \partial_u M+\frac{1}{4} \, D^AD^B(f \, \partial_u \, C_{AB})-\frac{1}{4} \, f \, D^AD^B \, \partial_u \, C_{AB},  \label{delfM}
\end{eqnarray}
and under superrotation  as
\begin{multline}
\delta_{\zeta} M= \zeta^z \, \partial_z \, M+\frac{3}{2} \, D_z\zeta^z \, M+\frac{1}{2} \, D_z\zeta^z \, u\partial_u M-\frac{1}{2} \, D_z^2\zeta^z \, u\partial_uU^z \\  
+\frac{1}{2}\zeta^z u\partial_uU_z 
+\frac{1}{8}\, D_z^3\zeta^z \, (u\partial_u+1) C^{zz}. \label{delzetaM}
\end{multline}

Similarly, angular momentum $L_A$ is defined by\footnote{By comparing transformation rules, it is found that $L_z$ is related to $N_z$ defined in \cite{BT2} by $L_z=-N_z-\frac{1}{32} D_z (C_{AB} C^{AB})$.}  
\begin{equation}
L_A \equiv  \frac{3}{2}W_A-\frac{1}{2} \, C_{AB}U^B-\frac{3}{8} \, D_A {H_A}^A+\frac{5}{32} \, D_A (C_{BC}C^{BC}) \label{LA}
\end{equation}
and transforms under supertranslation as 
\begin{multline}
\delta_f \, L_z= -3M \, \partial_z f+f \, \dot{L}_z -\frac{1}{4} \, D_z^3f \, C^{zz}-\frac{3}{4} \, D_z \, (D_zf \, D_z \, C^{zz})\\-\frac{3}{4} \, D_z(D^z)^2f \, C_{zz}
+\frac{3}{4} \, D_zf \, (D^z)^2 \, C_{zz}-\frac{1}{4}(D^z)^2f D_zC_{zz}\\+D_zf \, (-\frac{3}{8} \, C_{zz}\dot{C}^{zz}+\frac{1}{8}\dot{C}_{zz}C^{zz}).
\end{multline}
Its transformation rule under superrotation is given by  
\begin{eqnarray}
\delta_{\zeta} \, L_{z} &=& \frac{1}{2} \, u \, D_z\zeta^z \, \dot{L}_z+\zeta^z \, D_zL_z+2D_z\zeta^z \, L_z-\frac{3}{2} \, M \, D_z^2\zeta^z \nonumber \\&&-\frac{1}{2} \, \zeta^z \, H_{zz}-\frac{1}{8} \, u \, D_z^4\zeta^z \, C^{zz}+\frac{3}{8} \, u \, D_z^2\zeta^z \, (D^z)^2\, C_{zz} 
-\frac{3}{8}u\, D_z^3 \,\zeta^z \,  D_zC^{zz}\nonumber \\&&-\frac{3}{8} u D_z^2 \zeta^z D_z^2C^{zz}-\frac{3}{16}uD_z^2\zeta^zC_{zz}\dot{C}^{zz}+\frac{1}{16}uD_z^2\zeta^z\dot{C}_{zz}C^{zz}, \\
\delta_{\bar{\zeta}} \, L_z&=& \frac{1}{2} u\, D_{\bar{z}} \, \zeta^{\bar{z}} \, \dot{L}_z+\zeta^{\bar{z}} \, D_{\bar{z}}L_{z}-\frac{3}{2} \, D_{\bar{z}} \, \zeta^{\bar{z}} \, L_z-\frac{3}{2} \, h_{z\bar{z}} \, \zeta^{\bar{z}} \, M \nonumber \\ &&
+h_{z\bar{z}} \, \Big[\frac{1}{2} \,  ( h^{z\bar{z}})^2 \, H_{zz}D_{\bar{z}}^2\zeta^{\bar{z}}-\frac{3}{8} \, u \, \zeta^{\bar{z}} \, (D^z)^2C_{zz} -\frac{1}{16}u\, \zeta^{\bar{z}} \, C^{zz}\dot{C}_{zz}\nonumber \\
&& -\frac{1}{8}u\, D_{\bar{z}}^3\zeta^{\bar{z}} \, D^{\bar{z}} C^{\bar{z}\bar{z}}+\frac{3}{8}u \, \zeta^{\bar{z}} \, D_z^2 C^{zz}+\frac{3}{16}u \, \zeta^{\bar{z}} \, C_{zz}\dot{C}^{zz} \Big]. \label{delzbLz}
\end{eqnarray}

Due to the last terms in (\ref{delQf})-(\ref{delQeta}), 
 the variation equations are in general not integrable. 
One  prescription to ensure integrability will be to require $\dot{C}_{AB}=0$.  Then the surface charges become independent of $u$,  and conserved. In the next subsection this prescription will be studied. 

Another prescription will be to add  to  (\ref{delQf})-(\ref{delQeta})   
 an additional term such as 
\begin{equation}
\delta \, Q^{\text{additional}}[\xi]=\frac{-1}{32\pi G} \, \int_S d^2 z \, h_{z\bar{z}} \,  \dot{C}^{AB} \, \delta \,  C_{AB} \, \xi^{u} \label{extra}
\end{equation}
to drop the second terms in (\ref{delQf}) and (\ref{delQeta}) as was proposed in \cite{Wald}. By integrating the new variation equation $\delta \hat{Q}[\xi]= \delta Q[\xi]+\delta Q^{\text{additional}}[\xi]$, {\em i.e.},
\begin{eqnarray}
\delta \hat{Q}[f,0,0]&=& \frac{1}{16\pi G} \, \int_S d^2z \, h_{z\bar{z}} \,4 \,  f \, \delta M,  \label{delhatQf}\\
\delta \hat{Q}[0,\eta,0]&=&\frac{1}{16\pi G} \, \int_S d^2z \, h_{z\bar{z}} \, \eta^z \, \delta (-2L_z-2uD_z M), \label{delhatQeta}
\end{eqnarray} 
the surface charges $\hat{Q}[f,0,0]$ and $\delta Q[0,\eta,0]$ are obtained,
\begin{equation}
\hat{Q} [0,0,f] \equiv \frac{1}{16\pi G} \, \int d^2z \, h_{z\bar{z}}\,  4\, f \,  M,  \label{hQf}
\end{equation}
\begin{equation}
\hat{Q}[0,\eta,0] \equiv \frac{1}{16\pi G} \, \int d^2 z h_{z\bar{z}} \, \eta^z \, \Big[ -2 \, L_z-2u\partial_z M 
\Big].  \label{hQeta}
\end{equation}
These charges coincide with the \lq integrable part of the surface charge'  proposed in \cite{BT2}.\footnote{In eq (3.5) of \cite{BT2}, in order to define the algebra,  the right hand side of (\ref{QQdelQ}) is modified by decomposing the variation of surface charge $\delta \, Q[\xi]$ into an integrable part $\delta \, Q^{\text{int}}[\xi]$ and a non-integrable part $\delta Q^{\text{nonint}}[\xi]$, and replacing the non-integrable part  $\delta_{\xi'}\,  Q^{\text{nonint}}[\xi]$ by $\delta_{\xi}\,  Q^{\text{non}}[\xi']$. The left hand side is interpreted as $\{Q^{\text{int}}[\xi], Q^{\text{int}}[\xi']\}_D$.}
The algebra of these surface charges must be computed by using (\ref{QQdelQ}) with $Q[\xi]$ replaced by $\hat{Q}[\xi]$ and via the transformation rules (\ref{delfM})-(\ref{delzbLz}). 
 In this case, however,  it can be shown that the anti-symmetry property (\ref{antisym}) is not respected. 
For instance, by using (\ref{delfM}) for the supertranslation of $M$, the following Dirac bracket is obtained. 
\begin{eqnarray}
&&\{\hat{Q}[f,0,0], \hat{Q}[f',0,0]\}_D \equiv
\delta_{f'} \, \hat{Q}[f,0,0]  \nonumber \\
&&= \frac{1}{16\pi G} \, \int_S \, d^2z \, h_{z\bar{z}} \, \{ 4ff' \, \dot{M}-2 \, (D^A \, f)(D^B \, f') \, \dot{C}_{AB}-(f \, D^AD^B \, f') \, \dot{C}_{AB}\}  \nonumber \\
&&\neq -\{\hat{Q}[f',0,0], \hat{Q}[f,0,0]\}_D
\end{eqnarray}
Hence, clearly,  this prescription to add an additional term to the variation equation does not work for obtaining charges which generate algebra.

\subsection{Conserved surface charges and algebra}
\hspace{.5cm}
In this subsection, condition $\dot{C}_{AB}=0$ will be imposed on field $C_{AB}$.\footnote{ It is discussed in 
\cite{BT2} that, with the modification of the algebra as described in footnote 11, the standard bms$_4$ charges are conserved in the absence of news and  they generate a centerless algebra. In \cite{BTcurrent} a current algebra for spatial components in the absence of news is obtained.  } Then the variation equations (\ref{delQf})-(\ref{delQeta}) become integrable.The surface charge of supertranslation, $Q[f,0,0]=Q^{\text{transl}} [f]$, and that of superrotation, $Q[0,\eta,0]=Q^{\text{rot}}[\eta]$ read  
\begin{equation}
Q^{\text{transl}} [f] \equiv \frac{1}{16\pi G} \, \int d^2z \, h_{z\bar{z}}\,  4\, f \,  M \Big|_{\dot{C}_{AB}=0},  \label{Qtra} 
\end{equation}
\begin{equation}
Q^{\text{rot}}[\eta] \equiv \frac{1}{16\pi G} \, \int d^2 z h_{z\bar{z}} \, \eta^z \, \Big[ -2 \, L_z-2u\partial_z M 
\Big]\Big|_{\dot{C}_{AB}=0}, \label{Qrot}
\end{equation}
where condition $\dot{C}_{AB}=0$ is imposed. Except for this constraint these charges are the same as (\ref{hQf}) and (\ref{hQeta}). 
These charges are conserved, if $D_z^3\, \eta^z=0$. This is checked by using (\ref{dm}), (\ref{W}), (\ref{LA}).

By using definitions (\ref{Qtra})-(\ref{Qrot}) and results (\ref{tt})-(\ref{rbr}) and by setting $\dot{C}_{AB}=0$, the algebra of surface charges is expressed as follows. $\bar{Q}^{\text{rot}}[\bar{\eta}]$ is a complex conjugate of (\ref{Qrot}).
\begin{eqnarray}
 \Big\{ Q^{\text{transl}}[f_1], Q^{\text{transl}}[f_2] \Big\}_D = 0,
\label{tt2}
\end{eqnarray}
\begin{equation}
\ \Big\{ Q^{\text{rot}}[\eta], Q^{\text{transl}}[f] \Big\}_D  
= Q^{\text{transl}} [[\eta,f]] 
-\frac{1}{16\pi G} \, \int_S d^2z \, h_{z\bar{z}} \, \frac{1}{2} \, f \, D_z^3 \, \eta^z \, C^{zz}
,\label{rt2}
\end{equation}
\begin{multline}
\Big\{Q^{\text{rot}}[\eta], Q^{\text{rot}}[\chi]\Big\} 
= Q^{\text{rot}}[ [\eta,\chi] ]  \\
+\frac{1}{64\pi G} \, \int_S d^2 z h_{z\bar{z}} \, (D_z\eta^z \, D_z^3\chi^z-D_z\chi^z \, D_z^3 \, \eta^z) \, u \, C^{zz}, \label{rr2}
\end{multline}
\begin{multline}
\ \Big\{ \bar{Q}^{\text{rot}}[\bar{\chi}], Q^{\text{rot}}[\eta]\Big\}_D 
=\frac{1}{16 \pi G} \, \int_S d^2z h_{z\bar{z}} \, \Big[\frac{1}{4} u\, D_z^3\eta^z \, D_{\bar{z}} \chi^{\bar{z}}  \, C^{zz}-\frac{1}{4} u\, 
D_{\bar{z}}^3\chi^{\bar{z}} \, D_{z} \eta^z \, C^{\bar{z}\bar{z}}
        \Big].
\label{rbr2}
\end{multline}

Let us note that there are central extensions which depend on field $C_{AB}$. However, the condition $\partial_u \, C_{AB}=0$ contradicts with the transformation rule for $C_{AB}$  (\ref{delzCzz}), if $D_z^3 \zeta \neq 0$.  Hence BMS$_4$ algebra is not extended and the above central extensions all vanish. 
\begin{equation}
\ \{Q[\xi], Q[\xi']\}_D =Q[[\xi,\xi']]  \label{simple}
\end{equation}

The above algebra satisfies Jacobi identity:
$ \{Q[\xi_1], \{Q[\xi_2], Q[\xi_3]\}_D\}_D+\{Q[\xi_2], \{Q[\xi_3], Q[\xi_1]\}_D\}_D
 +\{Q[\xi_3], \{Q[\xi_1], Q[\xi_2]\}_D\}_D=0$.  
%\begin{eqnarray}
%&& \{Q[\xi_1], \{Q[\xi_2], Q[\xi_3]\}_D\}_D+\{Q[\xi_2], \{Q[\xi_3], Q[\xi_1]\}_D\}_D\nonumber \\
%&& \qquad \qquad +\{Q[\xi_3], \{Q[\xi_1], Q[\xi_2]\}_D\}_D=0.   \label{Jacobi}
%\end{eqnarray}
This is checked  via $
 \{Q[\xi_1], \{Q[\xi_2], Q[\xi_3]\}_D\}_D=- \delta_{\xi_1} \, \{Q[\xi_2], Q[\xi_3]\}_D$,
%\begin{equation}
%\ \{Q[\xi_1], \{Q[\xi_2], Q[\xi_3]\}_D\}_D=- \delta_{\xi_1} \, \{Q[\xi_2], %Q[\xi_3]\}_D, \label{QQQ}
%\end{equation} 
and the right hand side is evaluated by substituting variation of fields into the integrand of $\{Q[\xi_2], Q[\xi_3]\}_D$.

\section{Summary}
\hspace{5mm}
In this paper surface charges for supertranslation $Q^{\text{transl}}$ (\ref{Qtra}) and superrotations $Q^{\text{rot}}$ ($\bar{Q}^{\text{rot}}$) (\ref{Qrot}) on ${\cal I}$ and the algebra of these charges are studied  via Hamiltonian framework. 
If a field $C_{AB}$ is independent of $u$, these surface charges generate algebra (\ref{simple}) and the central extensions of the algebra vanish. 
It would be desirable, if appropriate modification of surface charges $Q[\xi]$ could be found which generate BMS$_4$ algebra also for the case $\dot{C}_{AB} \neq 0$ so that the right hand side of the algebra would coincide with the changes of  the surface charges $\delta_{\xi'} \, Q[\xi]$ under transformation. 

We obtained surface charges on the $u \geq 0$ part of  ${\cal I}^+$, and their Dirac bracket algebra (\ref{tt2})-(\ref{rbr2}). The charges are expressed in terms of radiation fields $C_{AB}$ in Bondi frame (\ref{Bondi frame}). In Bondi frame, however,  there is no horizen at $u=0$ and this is not a special point on ${\cal I}^+$. Hence it is expected that the above bracket algebra also holds on the $u <0$ part of ${\cal I}^+$.  On the other hand, since the space-like region ($r > t$) is foliated by time-like hypersurface, at present it is not possible to check this directly. This issue needs further investigation.

\newpage
\setcounter{section}{0}
\renewcommand{\thesection}{\Alph{section}}

%\section{}
%\hspace{5mm}

\end{document}